\documentclass[aps,prl,twocolumn,groupedaddress,floatfix,showpacs]{revtex4}
\usepackage{amssymb}
\usepackage{graphicx}
\usepackage{subfigure}	 	
\usepackage[english]{babel}
\usepackage{float}
\usepackage{color}

\begin{document}

\title{Strong asymmetry for surface modes in nonlinear lattices with long-range coupling}

\author{Alejandro J. Mart\'inez}
\affiliation{Departamento de F\'isica, Facultad de Ciencias, Universidad
de Chile, Santiago, Chile}
\affiliation{Center for Optics and Photonics, Universidad de Concepci\'on, Casilla 4016, Concepci\'on, Chile}

\author{Rodrigo A.\ Vicencio}
\affiliation{Departamento de F\'isica, Facultad de Ciencias, Universidad
de Chile, Santiago, Chile}
\affiliation{Center for Optics and Photonics, Universidad de Concepci\'on, Casilla 4016, Concepci\'on, Chile}

\author{Mario I. Molina}
\affiliation{Departamento de F\'isica, Facultad de Ciencias, Universidad
de Chile, Santiago, Chile}
\affiliation{Center for Optics and Photonics, Universidad de Concepci\'on, Casilla 4016, Concepci\'on, Chile}

\begin{abstract}

We analyze the formation of localized surface modes on a nonlinear cubic waveguide array 
in the presence of exponentially-decreasing long-range interactions. We find that the long-range coupling induces a strong asymmetry between the focusing and defocusing cases for
the topology of the surface modes and also for the minimum power needed to generate them. In particular, for the defocusing case, there is an upper power threshold for exciting staggered modes, which depends strongly on the long-range coupling strength. The power threshold for dynamical excitation of surface modes increase (decrease) with the strength of long-range coupling for the focusing (defocusing) cases. These effects seem to be generic for discrete lattices with long-range interactions.

\end{abstract}


\maketitle

Arrays of coupled optical waveguides and periodic photonic lattices
constitute a current area of intense research activity,
due to the rich physical phenomena that arise when combining discreteness,
periodicity, nonlinearity and surface effects~\cite{rep1}.
Besides the interest stemming from the creation and controlling
of the propagation of light beams for their potential use in multiport
switching and routing of signals for envisioned all-optical
devices,  ``discrete optics'' has also recently become one of the
favorite tools for direct observation of  phenomena associated with
discrete, periodic media, such as Bloch oscillations~\cite{bloch oscillations},
Anderson localization~\cite{Anderson}, discrete breathers and
solitons~\cite{breathers},  to name a few.

A substantial amount of work has been devoted to the case of weakly-coupled
nonlinear waveguide arrays, where the mode overlap between neighboring
guides is small,  and the nonlinearity is strictly local. Recent
experimental and theoretical work in realistic systems such as dipole-dipole
interactions in Bose-Einstein condensates (BEC)~\cite{BE dipole} and discrete light
localization in nematic liquid crystals~\cite{nematic}, has stimulated research
into the effects of nonlocal effects. In general, nonlocal
nonlinearity tends to stabilize several types of solitons, such as dark solitons
in 3D dipolar BEC~\cite{Nath}, chirp-imprinted spatial solitons in
nematic liquid crystals~\cite{nematic}, optical vortex solitons~\cite{Minzoni}, rotating dipole solitons~\cite{rotating dipole DS} and azimuthons~\cite{azimuthons}. The effect of long-range dispersive
interactions on the other hand, has received comparatively less attention.
The  effect of power-law dispersion on anharmonic chains~\cite{power law dispersion},
as well as the  inclusion of second-order coupling in optical waveguide
arrays~\cite{zigzag,szameit},  suggests the onset of bistable effects.
Although at first sight, the inclusion of long-range coupling would seem to lead to an 
increase of the power level needed to excite a localized mode~\cite{zigzag}, there are also some counterintuitive results for the case of a single nonlinear (cubic) defocusing impurity. There, a small addition of
coupling to second nearest-neighbors, actually {\em decreases} the power threshold for the generation of a localized mode~\cite{mm_prb}.

On the other hand, surface states have attracted considerable attention of the community during the last five years. Unlike the case of fundamental bulk modes, where there is no 
minimum power to excite them, for one-dimensional surface states there is a power threshold for their excitation. When only nearest-neighbors interactions are considered,
this power is independent of the sign of the nonlinearity~\cite{surface1d}.
\begin{figure}[h!]
\centering
\includegraphics[width=0.44\textwidth]{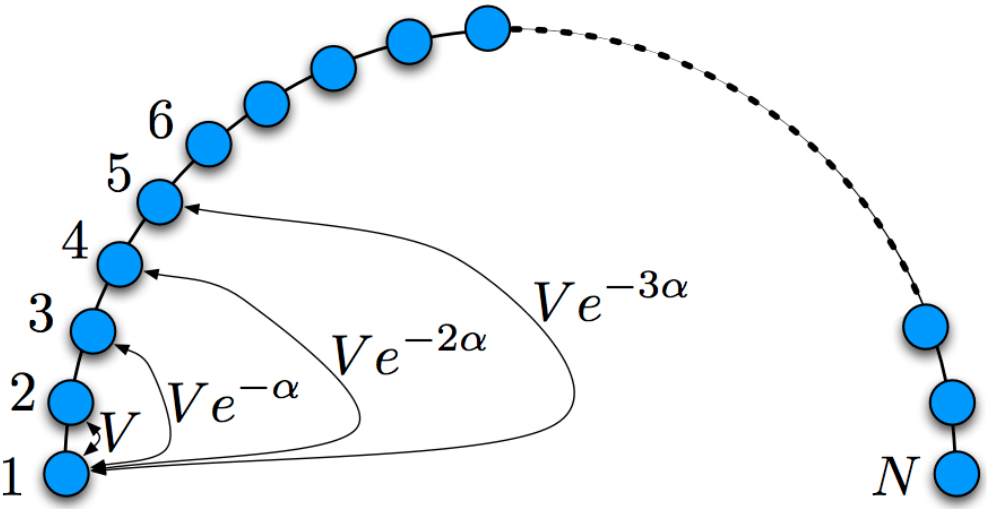}
\caption{$N$-sites waveguide array with long-range coupling.}
\label{fig1}
\end{figure}

In this work we examine the formation and stability of  localized surface modes in a nonlinear optical waveguide array with  realistic-looking long-range coupling (see Fig.\ref{fig1}). We find a striking asymmetry between the behavior of the focusing and defocusing cases, as the coupling range is varied.
Contrary to what occurs in a focusing case, for a defocusing nonlinearity an increase in coupling range actually reduces the amount of power needed to generate a surface localized stationary mode. This counterintuitive result also holds for the dynamical excitation of the surface mode from a narrow input beam. In addition, we found an \textit{upper threshold} for the excitation of staggered states, effect that could be experimentally observed in current zig-zag arrays~\cite{szameit}.

Let us consider a finite array of single-mode, nonlinear (Kerr) optical waveguides including higher-order coupling among sites. In the coupled-modes framework, the system is described by a discrete non-linear Schr\"{o}dinger (DNLS) equation:
\begin{equation}
i\frac{d u_n}{d z} + \sum_{m\neq n}V_{n,m}u_m + \gamma|u_n|^2u_n = 0,
\label{model}
\end{equation}
where $u_n$ is the amplitude of the waveguide mode in the {\it n}-th waveguide, $z$ is the propagation distance along the array, $\gamma$ is the nonlinear parameter, and the coefficient $V_{n,m}$ is the coupling between the {\it n}-th and {\it m}-th guides. To be consistent with coupled-mode approach, we will model $V_{n,m}$ as $V_{n,m}=V e^{-\alpha (|n-m|-1)}$, where $V$ is the usual coupling coefficient to nearest-neighbors and $\alpha>0$ is the strength for the long-range interaction. A large $\alpha$-value implies interaction with essentially one site (DNLS limit), while an small $\alpha$ increases the coupling range.
\begin{figure}[h]
\centering
\includegraphics[width=0.45\textwidth]{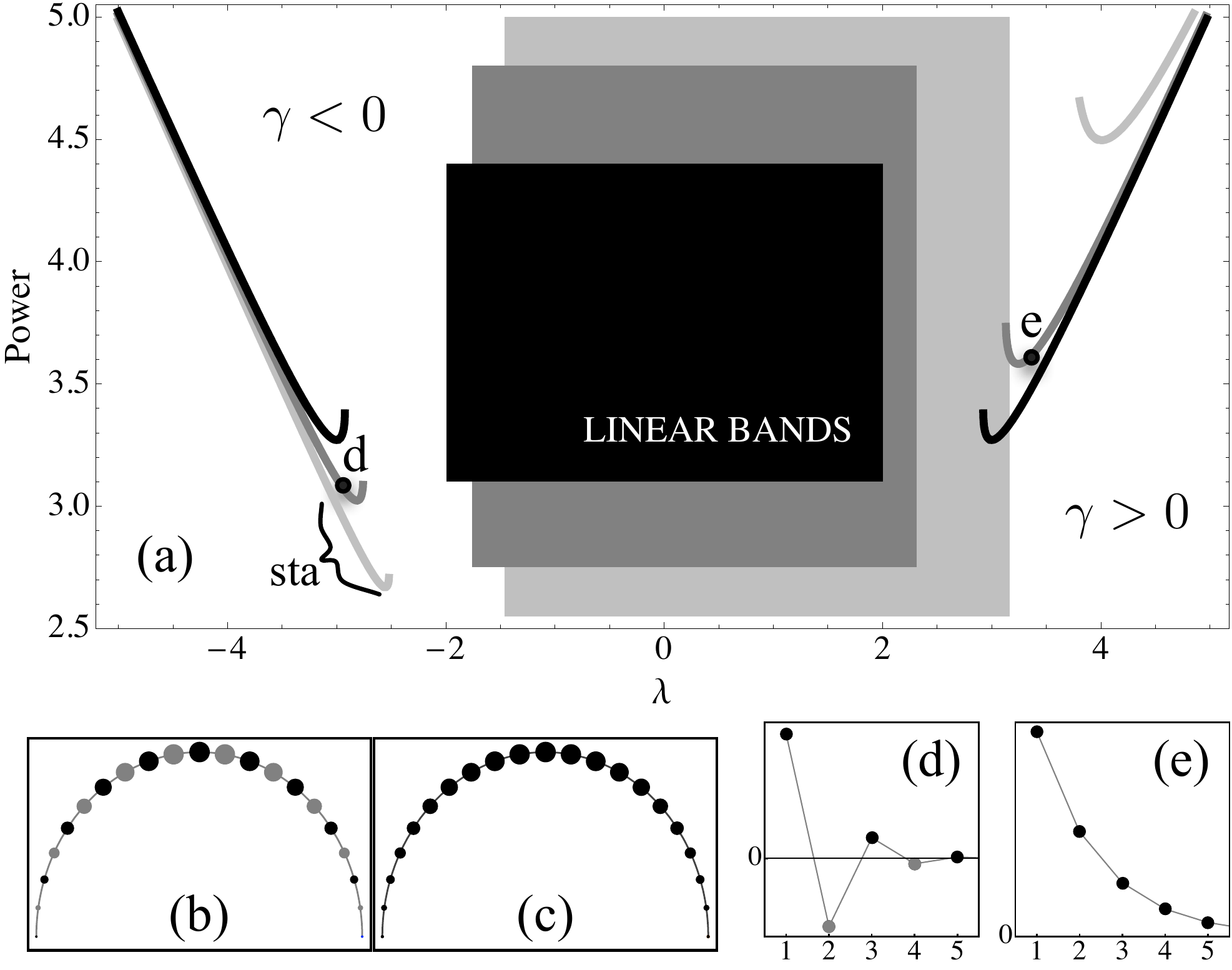}
\caption{(a) $P$ versus $\lambda$ diagrams (including linear bands) for $\alpha=6$ (black), $2$ (gray) and $1$ (light-gray), for focusing and defocusing cases. (b) and (c) Linear modes for $k=\pi$ and $k=0$, respectively. (d) and (e) Nonlinear surface modes for points marked in (a). $V=|\gamma|=1$ and $N=100$. Black (gray) points in profiles denote $u_{n}>0\    (<0)$.}
\label{fig2}
\end{figure}
The power, defined as $P=\sum_n |u_n|^2$, is a conserved quantity of model (\ref{model}) and we will use it to characterize nonlinear modes. We look for stationary solutions of the form $u_{n}(z) = u_{n} \exp(i\lambda z)$ of model (\ref{model}), obtaining:
\begin{equation}
\lambda u_{n} = \sum_{m\neq n} V e^{-\alpha(|n-m|-1)}\ u_m + \gamma u_n^3\ ,
\label{stati}
\end{equation}
where $u_n \in \Re$ and $\lambda$ is the propagation constant. 
To obtain the dispersion relation for linear plane waves, we set $\gamma = 0$ and insert a solution \mbox{$u_n = U \sin(k\hspace{0.05cm} n)$} in (\ref{stati}), getting
\begin{equation}
\lambda (k,\alpha)=V\left(\frac{e^{\alpha}\cos k - 1}{\cosh \alpha - \cos k}\right)\  ,
\label{dispersion}
\end{equation}
where $k$ is the transversal wave number. 
Fig.~\ref{fig2}(a) shows linear band regions for different values of $\alpha$. The edges of these bands are located at $\lambda_{min}\equiv\lambda (\pi,\alpha)=-2 V/[ 1+\exp(-\alpha)]$ and $\lambda_{max}\equiv\lambda (0,\alpha)=2 V/[ 1-\exp(-\alpha)]$, where the limit $N\rightarrow \infty$ has been assumed. The width $\lambda_{max}-\lambda_{min}= 4 V/[ 1-\exp(-2 \alpha)]$ increases as soon as coupling beyond nearest-neighbors is considered. As a consequence, the existence region for staggered solutions [$\lambda\in\{-\infty,\lambda_{min}\}$] increases with $\alpha$ while the corresponding region for unstaggered ones [$\lambda\in\{\lambda_{max},\infty\}$] decreases. Figures~\ref{fig2}(b) and (c) show profiles for $\lambda_{min}$ and $\lambda_{max}$ with a staggered [$(-1)^n u_n$] and an unstaggered ($u_n>0\ \forall n$) topologies, respectively.

Next, we compute nonlinear stationary surface solutions for focusing ($\gamma>0$) and defocusing ($\gamma<0$) nonlinearities by implementing a multi-dimensional Newton-Raphson method~\cite{surface1d}.
A linear stability analysis reveals that the Vakhitov-Kolokolov criterion still holds in the presence of long-range coupling, i.e., $\partial P/\partial \lambda>0$ implies stability. The P vs $\lambda$ curves for these modes show an important asymmetry between the focusing and defocusing cases [see Fig.\ref{fig2}(a)]:
In the short-range coupling case ($\alpha=6$, black curves), power thresholds ($P_{th}$) for positive and negative $\gamma$ are equal, like in a DNLS lattice \cite{surface1d}. However, when long-range coupling is relevant ($\alpha\ll 6$), $P_{th}$ increases as $\alpha$ decreases, in the focusing case. On the contrary, for $\gamma<0$ this threshold decreases when $\alpha$ decreases [see gray and light-gray curves in Fig.\ref{fig2}(a)].

An explanation for the $P_{th}$ asymmetry can be the following: We start from a surface profile, like the ones sketched in Fig.\ref{fig2}(d) and (e), with the general form: $u_{0}\{1,\epsilon,\beta,\xi,\ldots\}$ with $1>|\epsilon|>|\beta|>|\xi|>...$. By inserting this ansatz in (\ref{stati}) for $n=1$, we get: $\lambda=V(\epsilon+\beta e^{-\alpha}+\xi e^{-2\alpha}+...)+\gamma u_0^2$. Since discrete solitons exist outside of linear bands, fundamental localized solutions would - in principle - bifurcate exactly at the frontiers of these bands, depending on the sign of nonlinearity. Let us discuss first the \textit{unstaggered case} and try to get an estimate for $P_{th}$ in terms of $\alpha$. 
It is well known that when the solution approaches the linear band ($\lambda\rightarrow \lambda_{max}$), its power decreases and it becomes more and more extended (delocalized)~\cite{rep1, breathers}. This implies that (in such a limit) $\epsilon,\ \beta,\ \xi,...\rightarrow 1$ (this limit is exactly the opposite to the one occurring for high level of power, where solutions are extremely localized and $\epsilon,\ \beta,\ \xi,...\rightarrow 0$). Therefore, $1+e^{-\alpha}+e^{-2\alpha}+e^{-3\alpha}+...=1/(1-e^{-\alpha})$ implying that $\lambda\rightarrow V/(1-e^{-\alpha})+\gamma u_0^2$. However, this would imply that for $u_0\rightarrow 0$, $\lambda< \lambda_{max}$, which is certainly a contradiction because the fundamental unstaggered solution could originate from the top of the band but not inside of it. As a consequence, at least $\gamma u_0^2 \approx V/(1-e^{-\alpha})$. Since power is directly proportional to $u_0^{2}$, we obtain the estimate $P_{th}\sim 1/(1-e^{-\alpha})$. Thus, for $\gamma>0$, $P_{th}$ will be a decreasing function of $\alpha$, diverging at $\alpha =0$ and remaining finite at $\alpha\gg 0$. On the other hand, for $\gamma<0$, the situation is quite different. First of all, there is no a trivial transformation between unstaggered and staggered solutions as in the nearest-neighbor DNLS model. However, again, while the localized solution approaches the linear band ($\lambda\rightarrow \lambda_{min}$), its power decreases and it becomes more delocalized, but now the solution is staggered. That implies a sign difference between nearest-neighbor amplitudes in the same way as the fundamental linear mode located at $\lambda_{min}$ [see Fig.\ref{fig2}(b)]. Therefore $\epsilon, \ -\beta,\ \xi,...\rightarrow -1$. Now, we solve the sum: $1-e^{-\alpha}+e^{-2\alpha}-e^{-3\alpha}+...=1/(1+e^{-\alpha})$, implying that $\lambda\rightarrow -V/(1+e^{-\alpha})-|\gamma| u_0^2$. For $u_0\rightarrow 0$, $\lambda> \lambda_{min}$, i.e a contradiction. Again, at least $|\gamma| u_0^2 \approx V/(1+e^{-\alpha})$, so $P_{th}\sim 1/(1+e^{-\alpha})$. Thus, for $\gamma<0$,  $P_{th}$ is an increasing function of $\alpha$ with a minimum at $\alpha =0$. Our analytical estimates agree perfectly with the numerical behavior presented in Fig.\ref{fig2}(a).
\begin{figure}[h]
\centering
\includegraphics[width=0.48\textwidth]{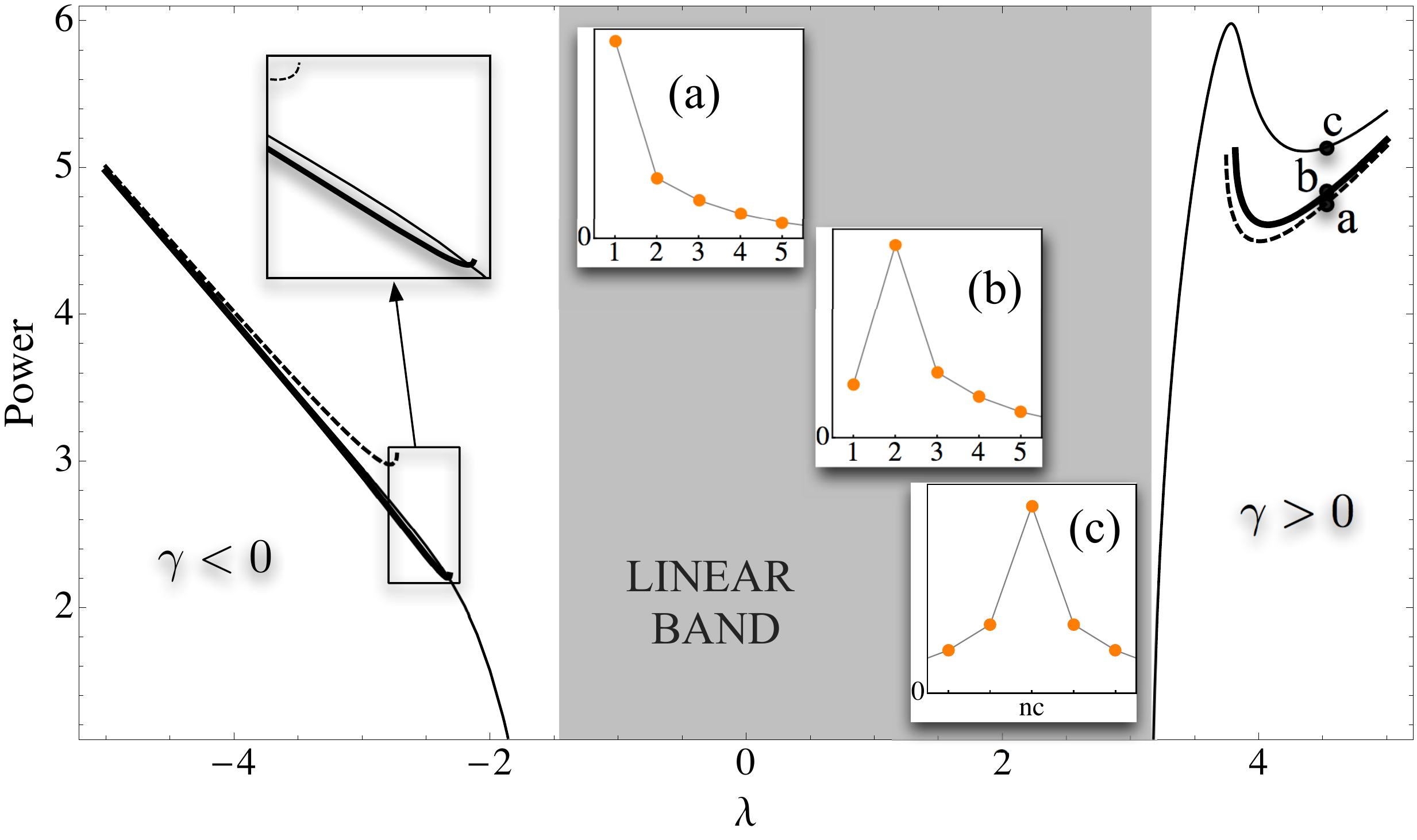}
\caption{$P$ versus $\lambda$ diagrams for modes centered at $n=1$ [dashed line,(a)], $n=2$ [thick line,(b)], and $n=n_c$ [thin line,(c)]. $V=|\gamma|=\alpha=1$ and $N=2n_c=100$.}
\label{fig3}
\end{figure}
%

We also computed localized solutions centered below the surface in order to detect the onset of the bulk phenomenology [see Fig.~\ref{fig3}]. For $\gamma>0$, the power as a function of $\lambda$ shows the onset of a bistable curve for $\alpha \lesssim 1.69$. This feature was observed before in the context of a zig-zag model~\cite{zigzag,szameit}, and seems to reflect an increase in effective dimensionality as soon as coupling beyond nearest-neighbors becomes important. The most salient feature is that in this case the threshold power to create a mode, behaves in a manner opposite to the usual DNLS. For example, for $\alpha=1$, Fig~\ref{fig3} shows that for $\gamma>0$, the minimum required power ($P_{th}$) for creating an unstaggered localized solution increases as the mode center is located away from the surface. In that sense, \textit{the system favors the localization of energy at the boundary for $\gamma>0$}, contrary to the usual 1D DNLS model~\cite{surface1d} (around $\alpha\approx 1.3$, the DNLS phenomenology transforms into the long-range one). On the other hand, the system asymmetry is manifest for $\gamma<0$; the $P_{th}$ for exciting a staggered localized mode decreases as the mode center is pushed away from the surface. Now, the system does not favor the generation of discrete surface solitons, as in the usual DNLS.
\begin{figure}[b]
\centering
\includegraphics[width=0.465\textwidth]{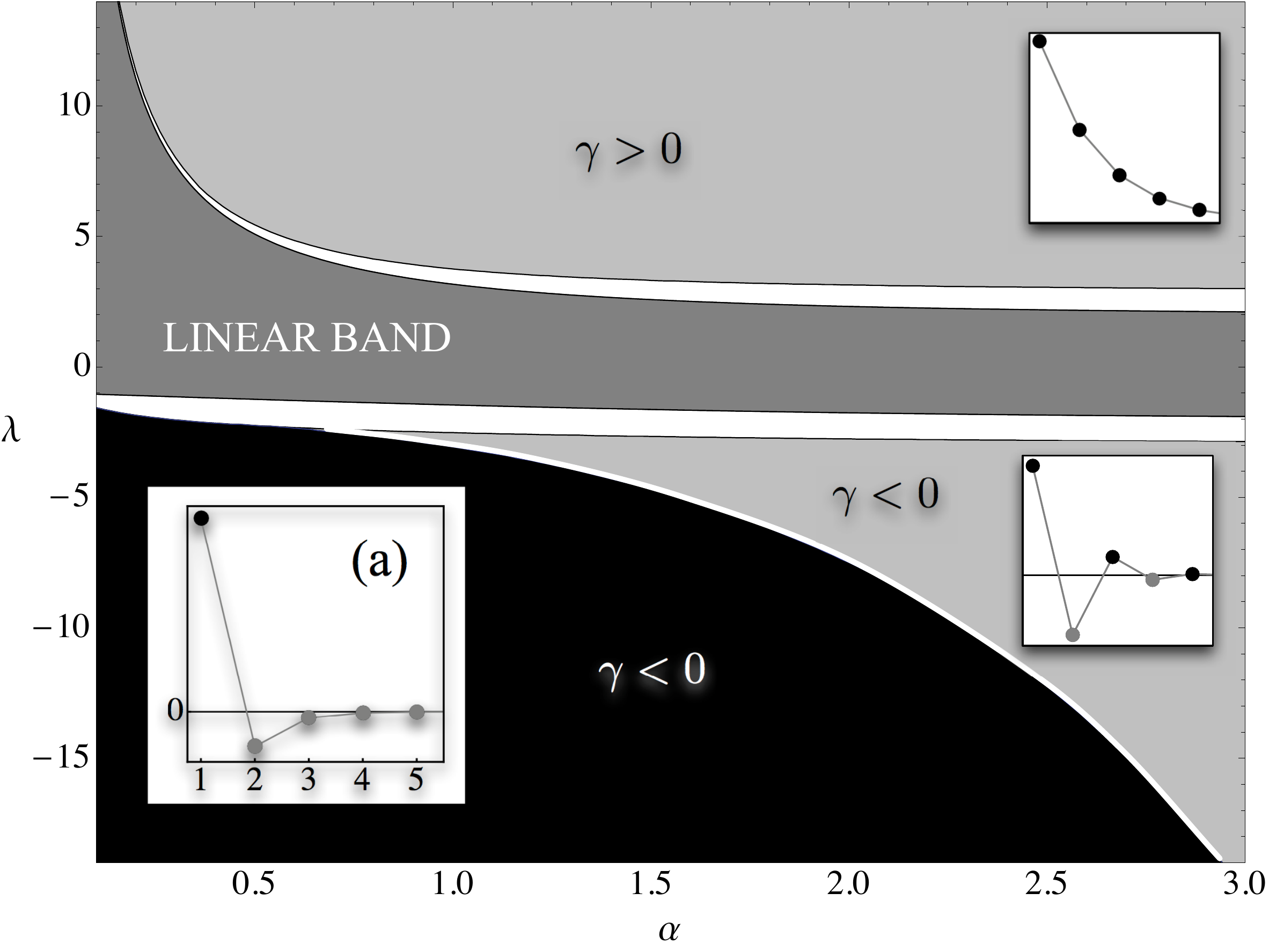}
\caption{Existence regions in $\alpha$-$\lambda$ space. The light-gray area corresponds to unstaggered ($\lambda>0$) and staggered ($\lambda<0$) solutions. The gray area represents the linear band while the white area denotes no solutions. The black region denotes solutions with no well-defined topology. Insets show examples of mode profiles in several regions.}
\label{fig4}
\end{figure}

Fundamental nonlinear modes are unstaggered for $\gamma>0$ and staggered for $\gamma<0$. As the power content of the mode is increased, we find that  unstaggered modes retain their character, as expected for high power solutions. However, contrary to what is expected, for staggered modes, a new power threshold ``$P_{up}$'' appears where the staggered topology of the mode is lost. Figure~\ref{fig4} shows the existence region for unstaggered-staggered solutions in $\alpha$-$\lambda$ space. We see that unstaggered solutions exist from some minimum $\lambda$ value (surface threshold) up to infinite. On the contrary, for $\gamma<0$ the staggered mode is well-defined from the surface threshold value - which depends weakly on $\alpha$ - up to a second $\lambda$ value (with power $P_{up}$), that increases monotonically with $\alpha$. Close to, but, beyond this second threshold, the mode is no longer staggered because although it retains some oscillations of the mode phases, it does not preserve a full staggered topology. As $\lambda$ decreases, the alternating phase topology is lost altogether. The white thick line separating the light-gray and black regions is a result of asking the solution if $u_{3}<0$, as an indicator of the change of topology in the central region. Fig.\ref{fig4}(a) shows a mode example where all lattice sites are negative excepting the first one, i.e the mode is - by definition - not staggered. In the numerical continuation there is no evidence of this change on topology [see ``sta'' in Fig.\ref{fig2}(a) for $\gamma<0$ and $\alpha=1$]. Curves are monotonous and the only way to observe this phenomenology is by taking a close look of phase structure.

We can use an strongly-localized mode approximation to give an explanation for this unexpected behavior occurring for $\gamma<0$. This approach is valid when the propagation constant is far from the linear band, where the mode can be approximated as $\{u_n\}=u_0 \{1,\epsilon,\beta,0,...\}$, and $1\gg|\epsilon|\gg|\beta|$. We concentrate the analysis in the parameter $\beta$, as an indicator of the long-range interaction effect. 
If we insert this ansatz in (\ref{stati}) and solve it for site $n=3$, we obtain: $\beta\approx V (\epsilon + e^{-\alpha})/(\lambda-\gamma u_0^2 \beta^2)$. From the anticontinuous limit, we know that high-power solutions consist essentially of one excited amplitude plus some exponentially small tails. Therefore, as a first approximation, $|\lambda| \gtrsim |\gamma| u_0^2$. If $\gamma>0$ also $\lambda,\epsilon>0$ [see Fig.\ref{fig2}(a) and (e)] implying that $\beta>0$ for any $\alpha$. This shows us that, for a focusing case, solutions preserve their phase in the whole range of parameters. On the contrary, when $\gamma<0$ also $\lambda,\epsilon<0$ [see Fig.\ref{fig2}(a) and (d)], therefore the sign of $\beta$ will depend on the balance $|\epsilon|-e^{-\alpha}$. For a fixed $\alpha$, this balance will be always negative for high-power solutions because $\epsilon\rightarrow 0$ (anticontinuous limit). Therefore, for a large $\alpha$ an upper power threshold is expected appearing at high frequencies; for smaller $\alpha$, this threshold is expected to occur closer to the band because, there, $\epsilon$ is also larger. This agrees perfectly with the thick white line of Fig.\ref{fig4}.
\begin{figure}[h]
\centering
\includegraphics[width=0.49\textwidth]{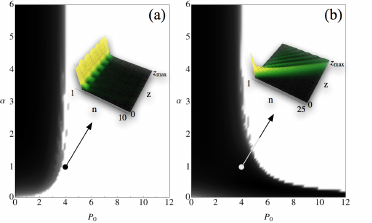}
\caption{Output power fraction in $P_0$-$\alpha$ space. Figures (a) and (b) correspond to $\gamma<0$ and $\gamma>0$. Dark and light regions denote $f=0$ and $f=1$, respectively. Insets: dynamical propagation for $\alpha=1$ and $P_0=4$.}
\label{fig5}
\end{figure}

We also looked into the effects of long-range coupling on the dynamical evolution of an initially localized input beam. We solved Eq.(1) numerically with initial condition $u_n(0)=\sqrt{P_0} \delta_{n,1}$, where $P_{0}$ is the input power. For a given $\alpha$ and $P_{0}$, we computed the space-averaged fraction of power remaining at the initial waveguide $f$, after a longitudinal propagation distance: $f =  (P_{0}\ z_{max})^{-1}  \int_{0}^{z_{max}} |u_{1}(z)|^{2} dz$. Results for $f$ as a function of $\alpha$ and $P_{0}$, are shown in Fig.\ref{fig5}, in the form of a density plot. Its most striking feature is the diametrically opposite selftrapping behavior between $\gamma>0$ and $\gamma<0$. While in the first case [Fig.\ref{fig5}(b)], an increase in coupling range (decrease $\alpha$) increases the threshold power for selftrapping, in the second case, a greater coupling range implies a smaller power threshold. This counterintuitive asymmetry becomes particularly strong around $\alpha\sim 1$ (see insets). These results are in complete agreement  with the ones  obtained for the stationary modes. 

Finally, we repeated all of the above studies on a simpler, but related model that is amenable to direct experimental probing: the zig-zag model~\cite{zigzag}, and have verified the strong asymmetry effects for the formation of localized surface modes  at both, the stationary and dynamics level. This opens the door to a direct experimental verification of these effects.

In conclusion, we have examined the formation of localized surface modes on a nonlinear waveguide array in the presence of realistic long-range interactions, and found a strong 
asymmetry between the focusing and defocusing cases for the mode topology and the minimum power to effect a localized surface mode.
We believe these effects are generic to discrete nonlinear systems with long-range 
coupling.

The authors are grateful to Y. S. Kivshar for useful discussions. This work was supported in part by FONDECYT Grants 1080374, 1070897, and Programa de Financiamiento Basal de CONICYT (FB0824/2008).

\end{document}